\documentclass{article} 
\usepackage{iclr2020_conference,times}
\usepackage{hyperref}

\usepackage{amsmath}
\usepackage{epstopdf}
\usepackage{graphicx}
\usepackage{multirow}

\graphicspath{{images/}}

\usepackage[justification=centering, labelfont=bf]{caption}
\usepackage{subfigure}
\usepackage{listings}
\usepackage{verbatim}
\usepackage{url}
\usepackage{footnote}
\usepackage{array}

\usepackage{tikz}

\let\origthelstnumber\thelstnumber
\makeatletter
\newcommand*\Suppressnumber{%
  \lst@AddToHook{OnNewLine}{%
    \let\thelstnumber\relax%
     \advance\c@lstnumber-\@ne\relax%
    }%
}

\newcommand*\Reactivatenumber[1]{%
  \setcounter{lstnumber}{\numexpr#1-1\relax}
  \lst@AddToHook{OnNewLine}{%
   \let\thelstnumber\origthelstnumber%
   \refstepcounter{lstnumber}
  }%
}

\def\setspacing#1{\renewcommand{\baselinestretch}{#1}\small\normalsize}
\setspacing{1.00}

\begin{document}


\title{Pipelined Training with Stale Weights of Deep Convolutional Neural Networks}
\author{Lifu Zhang \\
The Edward S.\ Rogers Sr.\ Department of \\ 
Electrical and Computer Engineering \\
University of Toronto \\
\texttt{lifu.zhang@mail.utoronto.ca} \\
\And Tarek S. Abdelrahman \\
The Edward S.\ Rogers Sr.\ Department of \\ 
Electrical and Computer Engineering \\
University of Toronto \\
\texttt{tsa@ece.utoronto.ca} \\
}

\maketitle

\begin{abstract}
The growth in the complexity of Convolutional Neural Networks (CNNs)
is increasing interest in partitioning a network across multiple accelerators
during training and pipelining the backpropagation computations over the
accelerators. Existing approaches avoid or limit the use of stale weights 
through techniques such as micro-batching or weight stashing. These techniques 
either underutilize of accelerators or increase memory 
footprint. We explore the impact of stale weights on the statistical efficiency 
and performance in a pipelined backpropagation scheme that maximizes accelerator 
utilization and keeps memory overhead modest. We use 4 CNNs (LeNet-5, AlexNet, 
VGG and ResNet) and show that when pipelining is limited to early layers in a 
network, training with stale weights converges and results in models with 
comparable inference accuracies to those resulting from non-pipelined training on MNIST 
and CIFAR-10 datasets; a drop in accuracy of 0.4\%, 4\%, 0.83\% and 1.45\% for the 4 networks, 
respectively. However, when pipelining is deeper in the network, inference accuracies drop 
significantly. We propose combining pipelined and non-pipelined training in 
a hybrid scheme to address this drop. We demonstrate the implementation and
performance of our pipelined backpropagation in {\em PyTorch} on 2 GPUs using
ResNet, achieving speedups of up to 1.8X over a 1-GPU baseline, with a 
small drop in inference accuracy.  
\end{abstract}


\section{Introduction}
\label{sec:introduction}

\begin{figure}[t]
    \centering
    \begin{minipage}{.5\textwidth}
        \centering
        \includegraphics[height=0.9in]{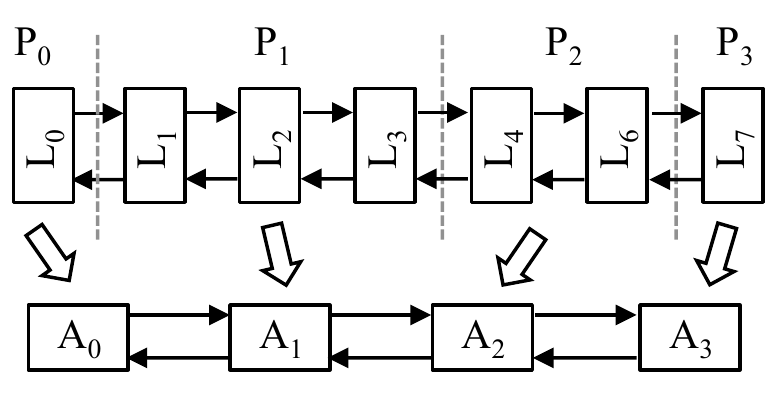}
        \vspace*{-0.1in}
        \caption{Partitioning of Layers}
        \label{fig:cnn-partition}
    \end{minipage}%
    \begin{minipage}{0.5\textwidth}
        \centering
        \includegraphics[height=0.9in]{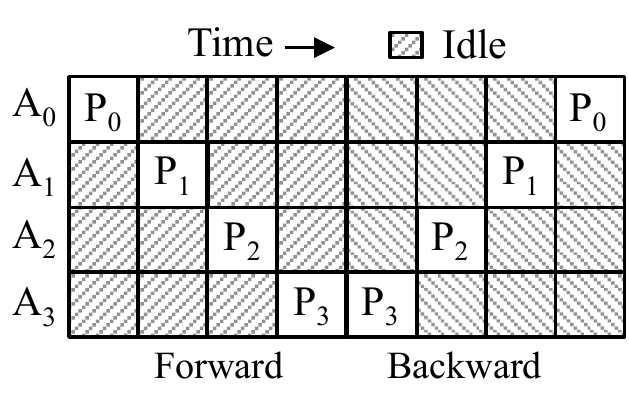}
        \vspace*{-0.1in}
        \caption{Schedule of Computations}
        \label{fig:spacetime-serial}
    \end{minipage}
\end{figure}

Modern Convolutional Neural Networks (CNNs) have grown in size and complexity to 
demand considerable memory and computational resources, particularly for training. 
This growth makes it sometimes difficult to train an entire network with a single 
accelerator~\citep{GPipe,PipeDream,Chen2012PipelinedBF}. Instead, the network is 
partitioned among multiple accelerators, typically by partitioning its layers 
among the available accelerators, as shown in Figure~\ref{fig:cnn-partition} 
for an example 8-layer network. The 8 layers are divided into 4 
computationally-balanced partitions, $P_0 ... P_3$ and each partition is mapped 
to one of the 4 accelerators, $A_0 ... A_3$. Each accelerator is responsible for 
the computations associated with the layers mapped to it.

However, the nature of the backpropagation algorithm used to train 
CNNs~\citep{backprop_algorithm} is that the computations of a layer are
performed only after the computations of the preceding layer in the forward 
pass of the algorithm and only after the computations of the succeeding 
layer in the backward pass. Further, the computations for one batch of input 
data are only performed after the computations of the preceding batch have 
updated the parameters (i.e., weights) of the network. These dependences 
underutilize the accelerators, as shown by the space-time diagram in 
Figure~\ref{fig:spacetime-serial}; only one accelerator can be 
active at any given point in time.

The underutilization of accelerators can be alleviated by {\em pipelining}
the computations of the backpropagation algorithm over the 
accelerators~\citep{GPipe,PipeDream,Chen2012PipelinedBF}.
That is, by overlapping the computations of different input data batches
using the multiple accelerators. However, pipelining causes an 
accelerator to potentially use weights that are yet to be updated
by an accelerator further down in the pipeline. The use of such {\em stale} 
weights can negatively affect the statistical efficiency of the network,
preventing the convergence of training or producing a model with lower 
inference accuracy.

Common wisdom is that the use of stale weights must either be avoided, 
e.g., with the use of micro-batches~\citep{GPipe}, be constrained to
ensure the consistency of the weights within an accelerator using 
stashing~\citep{PipeDream}, or by limiting the use of pipelining to 
very small networks~\citep{Mostafa}. However, these approaches either 
underutilize accelerators~\citep{GPipe} or inflate memory usage to 
stash multiple copies of weights~\citep{PipeDream}.

In this paper we question this common wisdom and explore pipelining 
that allows for the full utilization of accelerators while using 
stale weights. This results in a pipelining scheme that, compared
to existing schemes, is simpler to implement, fully utilizes the
accelerators and has lower memory overhead. We evaluate this 
pipelining scheme using 4 CNNs: LeNet-5 (trained on MNIST), 
AlexNet, VGG and ResNet (all trained on CIFAR-10). 
We analyze the impact of weight staleness and show that if pipelining 
is limited to early layers in the network, training does converge and 
the quality of the resulting models is comparable to that of models 
obtained with non-pipelined training. For the 4 networks, the drop
in accuracy is 0.4\%, 4\%, 0.83\% and 1.45\%, respectively.
However, inference accuracies drop significantly when the pipelining 
is deeper in the network. While this is not a limitation since the 
bulk of computations that can benefit from pipelining are in the 
early convolutional layers, we address this through a hybrid scheme 
that combines pipelined and non-pipelined training to maintain 
inference accuracy while still delivering performance improvement. 
Evaluation shows that our pipelined training delivers a speedup 
of up to 1.8X on a 2-GPU system. 

The remainder of this paper is organized as follows. Section~\ref{sec:backprop} 
briefly describes the backpropagation for training of CNNs. Section~\ref{sec:pipelined}
details our pipelining scheme. Section~\ref{sec:hybrid} describes how 
non-pipelined and pipelined backpropagation are combined. Section~\ref{sec:implementation} 
highlights some of the implementation details. Experimental evaluation is presented in 
Section~\ref{sec:evaluation}. Related work is reviewed in Section~\ref{sec:related}. Finally, 
Section~\ref{sec:conclusions} gives concluding remarks and directions for future work.

\newcommand{\given}{\,|\,}
\newcommand{\R}{\mathbb{R}}
\newcommand{\E}{\mathbb{E}}
\newcommand{\var}{\text{var}}
\newcommand{\cov}{\text{cov}}
\newcommand{\trans}{\mathsf{T}}
\newcommand{\bx}{\mathbf{x}}
\newcommand{\by}{\mathbf{y}}
\newcommand{\bc}{\mathbf{c}}
\newcommand{\bt}{\mathbf{t}}
\newcommand{\bw}{\mathbf{w}}
\newcommand{\bW}{\mathbf{W}}
\newcommand{\bs}{\mathbf{s}}
\newcommand{\bX}{\mathbf{X}}
\newcommand{\distNorm}{\mathcal{N}}
\newcommand{\bzero}{\mathbf{0}}
\newcommand{\btheta}{\boldsymbol{\theta}}
\newcommand{\bdelta}{\boldsymbol{\delta}}
\newcommand{\bpi}{\boldsymbol{\pi}}
\newcommand{\bmu}{\boldsymbol{\mu}}
\newcommand{\bsigma}{\boldsymbol{\sigma}}
\newcommand{\bphi}{\boldsymbol{\phi}}
\newcommand{\ident}{\mathbb{I}}
\newcommand{\N}{\mathcal{N}}
\newcommand{\ep}{\varepsilon}
\newcommand{\Dir}{\text{Dirichlet}}

\section{The Backpropagation Algorithm}
\label{sec:backprop}

The {\em backpropagation} algorithm~\citep{backprop_algorithm} consists 
of two passes: a {\em forward} pass that calculates the output error 
and a {\em backward} pass that calculates the error gradients and updates 
the weights of the network. The two passes are performed for input data 
one {\em mini-batch} at a time. 



In the forward pass, a mini-batch is fed into the network, propagating 
from the first to the last layer. At each layer $l$, the activations 
of the layer, denoted by $\bx^{(l)}$, are computed using the weights 
of the layer, denoted by $\bW^{(l)}$. When the output of the network
(layer $L$) $\bx^{(L)}$ is produced, it is used with the true data 
label to obtain a training error $e$ for the mini-batch.

In the backward pass, the error $e$ is propagated from the last to the 
first layer. 
The error gradients with respect to pre-activations of layer $l$, denoted by
$\bdelta^{(l)}$, are calculated. Further, the error gradients with respect to weights 
of layer $l$, $\frac{\partial{e}}{\partial{\bW^{(l)}}}$, are computed using 
the activations from layer $l-1$ (i.e., $\bx^{(l-1)}$) and $\bdelta^{(l)}$.  
Subsequently, $\bdelta^{(l)}$ is used to calculate the $\bdelta^{(l-1)}$. When 
$\frac{\partial{e}}{\partial{\bW^{(l)}}}$ is computed for every layer, the weights 
are updated using the error gradients.

In the forward pass, the activations of the layer $l$, $\bx^{(l)}$, cannot 
be computed until the activations of the previous layers, i.e., $\bx^{(l-1)}$, 
are computed.  In backward pass, $\frac{\partial{e}}{\partial{\bW^{(l)}}}$ 
can only be computed once $\bx^{(l-1)}$ and $\bdelta^{(l)}$ have been computed. 
Moreover, $\bdelta^{(l)}$ depends on $\bdelta^{(l+1)}$. Finally, for a 
given mini-batch the backward pass cannot be started until the forward
pass is completed and the error $e$ has been determined. 

The above dependences ensure that the weights of the layers are updated 
using the activations and error gradients calculated from the {\em same} 
batch of training data in one iteration of the backpropagation algorithm. 
Only when the weights are updated is the next batch of training data fed 
into the network. These dependences limit parallelism when a network is 
partitioned across multiple accelerators and allow only one accelerator to
be active at any point. This results in under-utilization of the
accelerators. It is this limitation that pipelining addresses.

\section{Pipelined Backpropagation}
\label{sec:pipelined}

\begin{figure}[tp]
  \begin{center}
    \includegraphics[width=.7\linewidth]{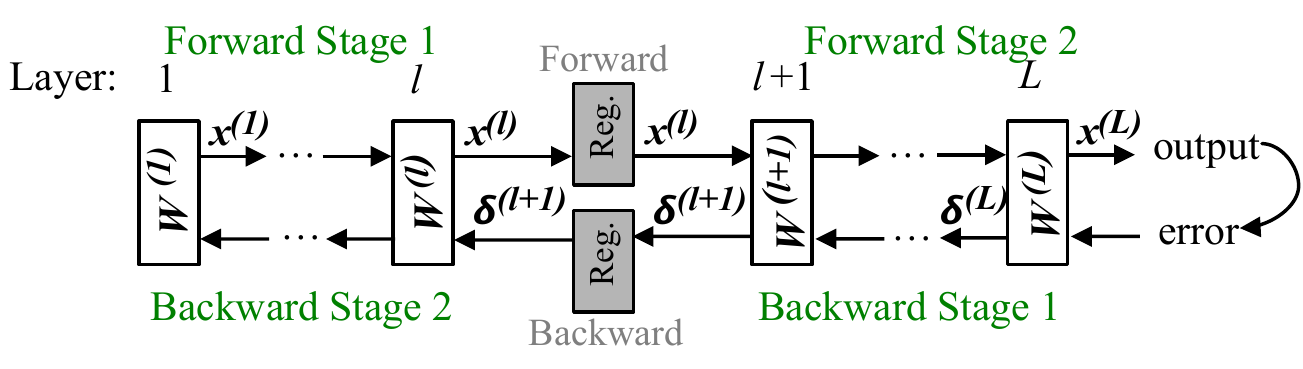}
    \vspace*{-0.1in}
    \caption{Pipelined Backpropagation Algorithm}
    \label{fig:pipelinedbackprop}
  \end{center}
\end{figure}

We illustrate our pipelined backpropagation implementation with the $L$ layer 
network shown in Figure~\ref{fig:pipelinedbackprop}, using conceptual pipeline 
registers. Two registers are inserted between layers $l$ and $l+1$; one 
register for the forward pass and a second for the backward pass. The 
forward register stores the activations of layer $l$ (${\bf x}^{(l)}$). 
The backward register stores the gradients ${{\bf \delta}^{(l+1)}}$ of layer $l+1$. 
This defines a {\em 4-stage} pipelined backpropagation. The forward pass 
for layers $1$ to $l$ forms forward stage ${\rm FS}_1$. The forward pass for layers 
$l+1$ to $L$ form forward stage ${\rm FS}_2$.  Similarly, the backwards pass for layers 
$l+1$ to $L$ and $1$ to $l$ form backward stages ${\rm BKS}_1$ and ${\rm BKS}_2$ 
respectively.

\begin{figure}[tp]
  \begin{center}
    \includegraphics[width=.55\linewidth]{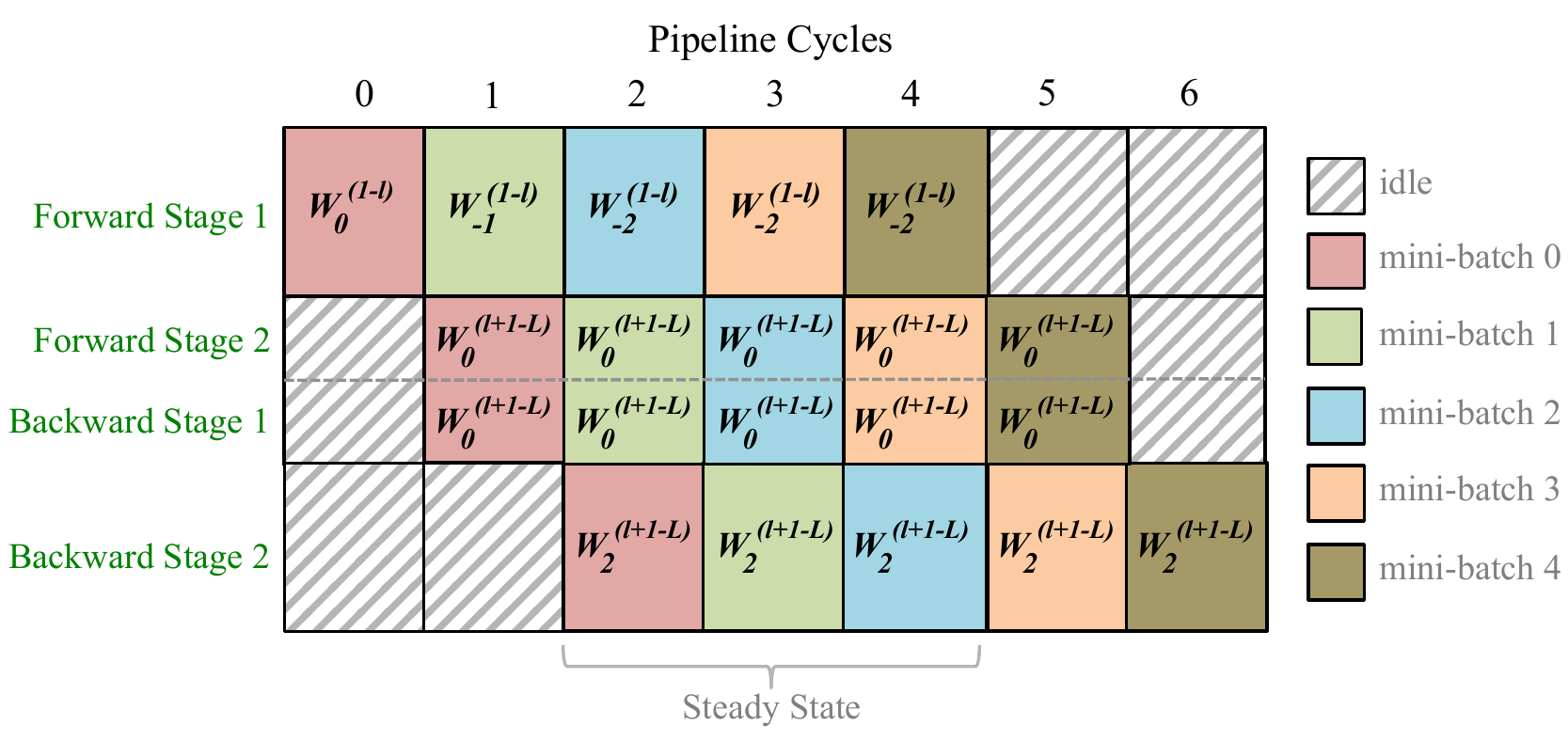}
    \vspace*{-0.1in}
    \caption{An Illustration of Computations of The Pipelined Backpropagation for Each Cycle}
    \label{fig:cycles}
  \end{center}
\end{figure}

The forward and backward stages are executed in a pipelined fashion on 3
accelerators: one for ${\rm FS}_1$, one for both ${\rm FS}_2$ and ${\rm BKS}_1$, 
and one for ${\rm BKS}_2$\footnote{We combine  ${\rm FS}_1$ and ${\rm BKS}_1$ 
on the same accelerator to reduce weight staleness.}.
In cycle~0, mini-batch 0 is fed to ${\rm FS}_1$. The computations of the 
forward pass are done as in the traditional non-pipelined implementation.  
In cycle~1, layer $l$ activations ${\bf x}^{(l)}$ are fed to ${\rm FS}_2$ and mini-batch 1 
is fed to ${\rm FS}_1$. In cycle~2, the error for mini-batch~0 computed in 
${\rm FS}_2$ is directly fed to ${\rm BKS}_1$, the activations of layer $l$ ${\bf x}^{(l)}$
are forwarded to ${\rm FS}_2$ and mini-batch~2 is fed to ${\rm FS}_1$.
This pipelined execution is illustrated by the space-time diagram in 
Figure~\ref{fig:cycles} for 5 mini-batches. The figure depicts the mini-batch 
processed by each accelerator cycles 0 to 6. At steady state, all the 
accelerators are active in each cycle of execution. 

The above pipelining scheme utilizes weights in ${\rm FS}_1$ that
are yet to be updated by the errors calculated by ${\rm FS}_2$ and ${\rm BKS}_1$.
At steady state, the activations of a mini-batch in ${\rm FS}_1$ are calculated 
using weights that are 2 execution cycles old, or 2 cycles {\em stale}. This is 
reflected in Figure~\ref{fig:cycles} by indicating the weights used by each 
forward stage and the weights updated by each backward stage. The weights 
of a forward stage are subscripted by how stale they are (-ve subscripts). 
Similarly, the weights updated by a backward stage are subscripted by how 
delayed they are (+ve subscripts).

Further, since the updates of the weights by ${\rm BKS}_2$ requires activations
calculated for the same mini-batch in ${\rm FS}_1$ for all layers in the stage, 
it is necessary to save these activations until the error gradients with respect 
to the weights are calculated by ${\rm BKS}_2$. Only when the weights are updated 
using the gradients can these activations be discarded. 

In the general case, we use $K$ pairs of pipeline registers (each pair consisting of
a forward register and a backward register) inserted between the layers of the
network.  We describe the placement of the register pairs by the {\em Pipeline 
Placement Vector}, ${\rm PPV} = (p_{1},p_{2},...,p_{K})$, where $p_i$ represents the layer 
number after which a pipeline register pair is inserted. Such a placement creates
$(K+1)$ forward stages, labeled ${\rm FS}_i, i = 1, 2, ... , K+1$ and $(K+1)$ backward
stages, labeled  ${\rm BKS}_i, i = 1, 2, ... , K+1$. Forward stage ${\rm FS}_i$ and 
backward stage $BKS_{K-i+2}$ correspond to the same set of layers. Specifically,
stage ${\rm FS}_i$ contains layers $p_{i}+1$ to $p_{i+1}$, inclusive.
We assign each forward stage and each backward stage to an accelerator, with the 
exception of the ${\rm FS}_{K+1}$ and backward stage ${\rm BKS}_1$, which are assigned to 
the same accelerator to reduce weight staleness by an execution cycle. In total 
$2K+1$ accelerators are used.

We quantify weight staleness as follows. A forward stage ${\rm FS}_i$ and backward 
stage ${\rm BKS}_{K-i+2}$ use the same weights that are $2(K-i+1)$ cycles old. 
A forward stage ${\rm FS}_i$ must store the activations of all layer in the 
stage for all $2(K-i+1)$ cycles which are used for the corresponding backward 
stage ${\rm BKS}_{K-i+2}$. Thus, we define the {\em Degree of Staleness} as $2(K-i+1)$, and
these saved activations as {\em intermediate activations}.
For each pair of stages ${\rm FS}_{i}$ and ${\rm BKS}_{K-i+2}$, let there be $N_{i}$ 
weights in their corresponding layers. 
The layers before the last pipeline register pairs always use stale weights. 
Thus, we define {\em Percentage of Stale Weight} as $(\sum_{i=1}^{K} N_{i})/(\sum_{i=1}^{K+1} N_{i})$. 

On the one hand, the above pipelined execution allows a potential
speedup of $2K+1$ over the non-pipelined implementation, keeping all the
accelerators active at steady state. On the other hand, the use of stale 
weights may prevent training convergence or may result in a model that has 
an inferior inference accuracy. Further, it requires an increase in storage 
for activations. Our goal is to assess the benefit of this pipelined 
execution and the impact of its down sides.

\section{Hybrid Pipelined/Non-Pipelined Backpropagation}
\label{sec:hybrid}

{\em Hybrid} training combines pipelined training with non-pipelined 
training. We start with pipelined training and after a number 
of iterations, we switch to non-pipelined training. This can 
address drops in inference accuracy of resulting models 
because of weight staleness, but it reduces the performance
benefit since during non-pipelined training, the accelerators are
under-utilized. 

The extent of the speedup obtained by hybrid training with a 
given number of accelerators is determined by the number of
iterations used for pipelined and non-pipelined training. Assume
that $n_{np}$ iterations are used to reach the best inference 
accuracy for non-pipelined training, and that in hybrid training,
$n_p$ iterations ($n_p \leq n_{np}$) are pipelined followed by 
$n_{np} - n_p$ iterations of non-pipelined training to reach the 
same inference accuracy as non-pipelined training. The speedup 
of hybrid training with respect to the non-pipelined training 
with $2K+1$ accelerators is $n_{np}/(n_p/(2K+1)+(n_{np}-n_p))$. 
For large $K$, then using Amdahl's law, the speedup approaches an 
upper bound of $n_{np}/(n_{np}-n_p)$.

\section{Implementation}
\label{sec:implementation}

We implement pipelined training in two ways: {\em simulated} in 
\emph{Caffe}~\citep{Caffe}, where the whole training process is 
performed on one process with no parallelism, and {\em actual}
with parallelism across accelerators in \emph{PyTorch}~\citep{PyTorch}. 
The simulated execution is used to analyze statistical convergence, 
inference accuracy and impact of weight staleness unconstrained
by parallelism and communication overhead. The actual execution is
used to report performance and {\em PyTorch} is used instead of
{\em Caffe} to leverage its support for collective communication 
protocols and its flexibility in partitioning a network across
multiple accelerators. Both {\em Caffe} and {\em PyTorch} 
have no support for pipelined training. Thus both were extended 
to provide such support.

We develop a custom \emph{Caffe} layer in Python,
which we call a Pipeline Manager Layer (PML), to facilitate 
the simulated pipelining. 
During the forward pass, a PML registers the input from a 
previous layer and passes the activation 
to the next layer. It also saves the activations for the 
layers connected to it to be used in the backward pass. During the
backward pass, a PML passes the appropriate error gradients. It 
uses the corresponding activations saved during the forward pass 
to update weights and generate error gradients for the previous 
stage, using existing weight update mechanisms in \emph{Caffe}.

To implement actual hardware-accelerated pipelined training, 
we partition the network onto different accelerators (GPUs), each 
running its own process. Asynchronous sends and receives are 
used for data transfers, but all communication must go 
through the host CPU, since point-to-point communication between 
accelerators is not supported in {\em PyTorch}. This increases 
communication overhead. Similar to the PMLs in {\em Caffe}, the 
activations computed on one GPU are copied to the next GPU 
(via the CPU) in the forward pass and the error gradients are 
sent (again via the CPU) to the preceding GPU during the 
backward pass. The GPUs are running concurrently, achieving 
pipeline parallelism.


\section{Evaluation}
\label{sec:evaluation}


\subsection{Setup, Methodology and Metrics}

Simulated pipelining is evaluated on a machine with one Nvidia
GTX1060 GPU with 6~GB of memory and an Intel~i9-7940X CPU with 64~GB 
of RAM. The performance of actual pipelining is evaluated using 
two Nvidia GTX1060 GPUs, each with 6~GB of memory, hosted in 
an Intel~i7-9700K machine with 32~GB of RAM.

We use four CNNs in our evaluation:  LeNet-5~\citep{LeNet-5} trained on 
MNIST~\citep{MNIST}, AlexNet~\citep{AlexNet}, VGG-16~\citep{VGG} and 
ResNet~\citep{ResNet}, all trained on CIFAR-10~\citep{CIFAR-10}.
For ResNet, we experiment with different depths: 20, 56, 110, 
224 and 362. We train these CNNs mostly following their original 
setting~\citep{LeNet-5}~\citep{AlexNet}~\citep{VGG}~\citep{ResNet} 
with minor variations to the hyperparameters, as described in 
Appendix~\ref{app:hyper}.

We evaluate the effectiveness of pipelined training in terms of its training
convergence and its {\em Top-1} inference accuracy, compared to those of the
non-pipelined training. We use the {\em speedup} to evaluate performance 
improvements. It is defined as the ratio of the training time of the 
non-pipelined implementation on single communication-free GPU to the 
training time of the pipelined training.

\subsection{Training Convergence and Inference Accuracy}

\begin{figure*}[t!]
    \centering
    \begin{subfigure}[LeNet-5 on MNIST]{
         \label{fig:lenet_pipelined}
                \includegraphics[width=.42\linewidth]{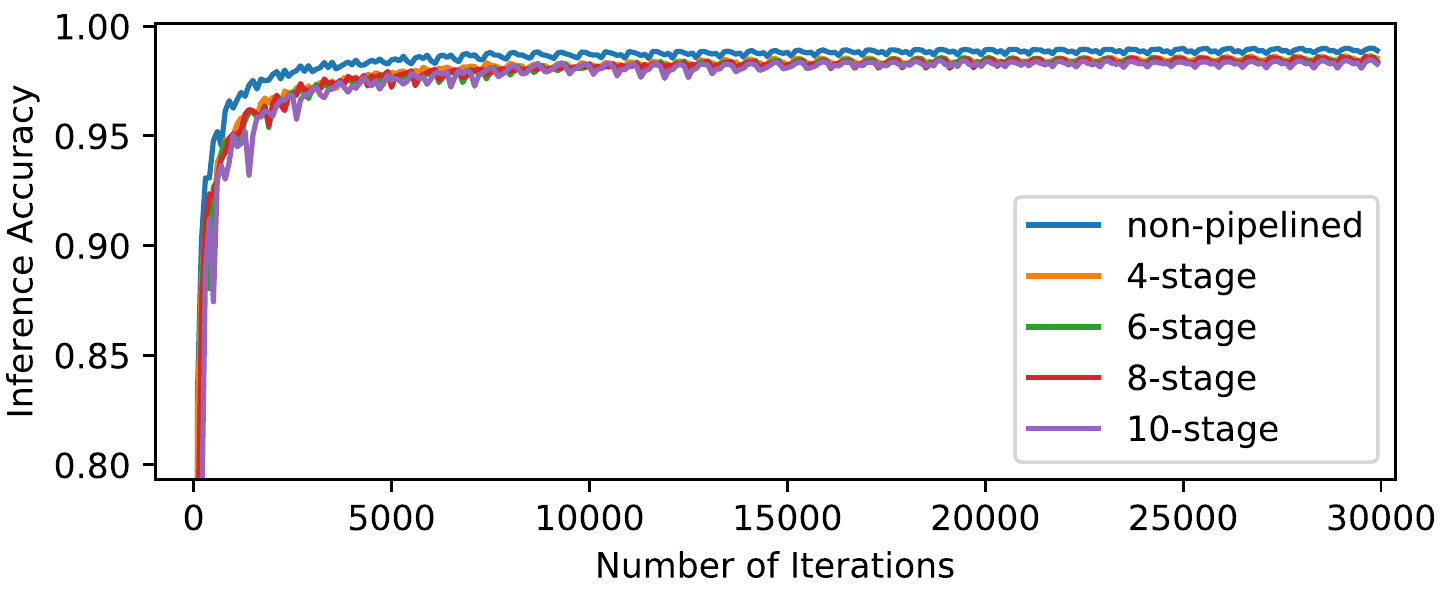}}
          \end{subfigure}
                \qquad
    \begin{subfigure}[AlexNet on CIFAR-10]{
        \label{fig:alexnet_pipelined}
        \includegraphics[width=.4\linewidth]{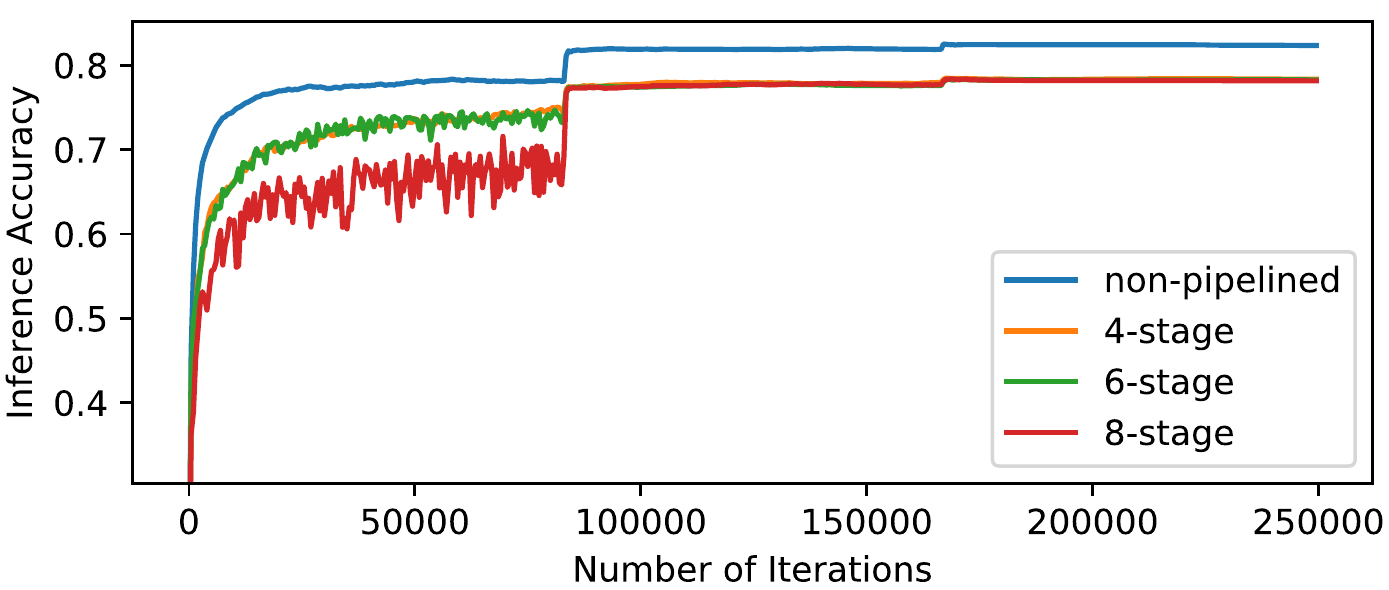}}
          \end{subfigure}
    \begin{subfigure}[VGG-16 on CIFAR-10]{
        \label{fig:vgg_pipelined}
        \includegraphics[width=.4\linewidth]{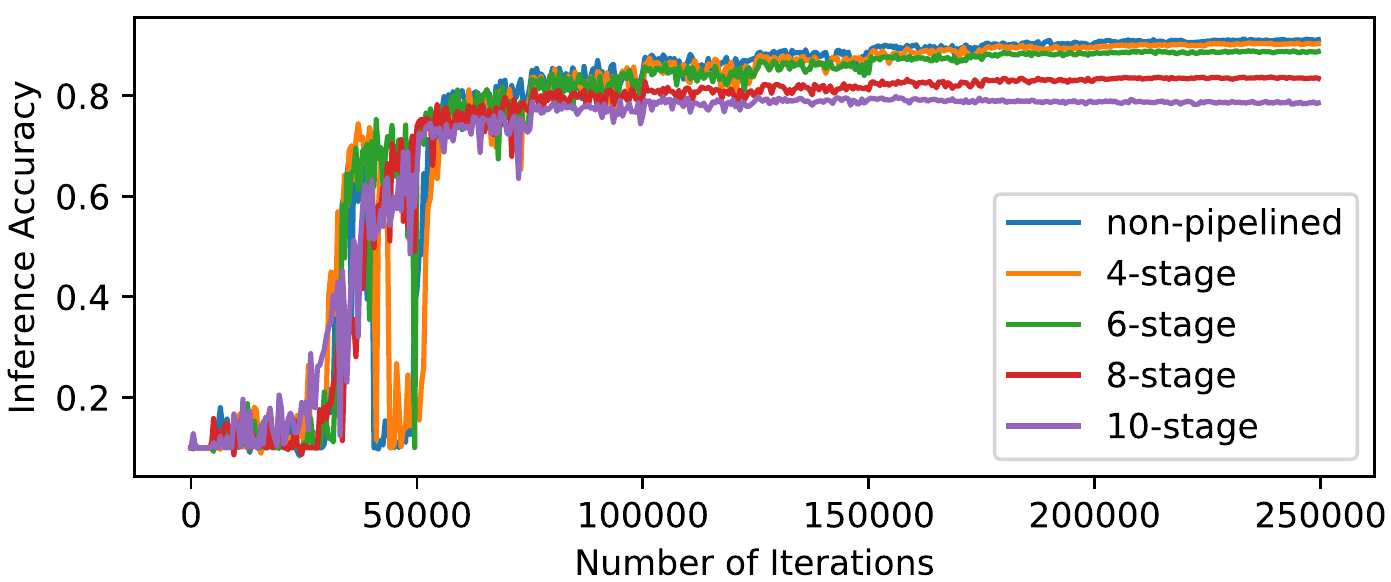}}
          \end{subfigure}
                \qquad
    \begin{subfigure}[ResNet-20 on CIFAR-10]{
        \label{fig:resnet_pipelined}
        \includegraphics[width=.4\linewidth]{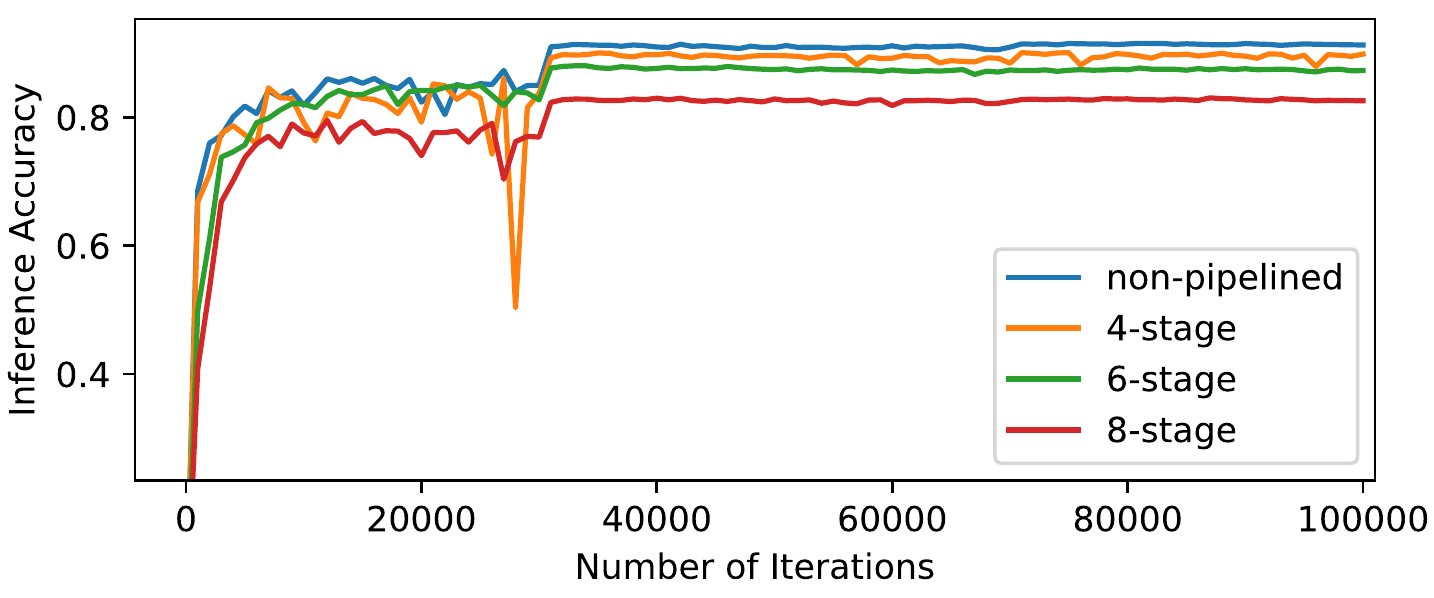}}
          \end{subfigure}
     \vspace*{-0.1in}
    \caption{Inference Accuracy Curves for Simulated Pipelined Training}
    \label{fig:pipelined_accuracy_curves}
    \end{figure*}

Figure~\ref{fig:pipelined_accuracy_curves} shows the improvements in the 
inference accuracies for both pipelined and non-pipelined training as a 
function of the number of training iterations (each iteration corresponds
to a mini-batch). The pipelined training is done using 4, 6, 8 and
10 stages. Table~\ref{table:pipeline_schemes} shows where the registers
are inserted in the networks using their ${\rm PPV}$ defined in 
Section~\ref{sec:pipelined}.
Figure~\ref{fig:pipelined_accuracy_curves} shows that for all the networks, 
both pipelined and non-pipelined training have similar convergence patterns. 
They converge in more or less the same number of iterations for
a given number of pipeline stages, albeit to different inference 
accuracies. This indicates that our approach to pipelined training
with stale weights does converge, similar to non-pipelined training.
 
\begin{table*}[t]
    \begin{center}
    \begin{tabular}{| c | c | c | c | c | c |}
    \hline
    CNN & Number of Layers & 4-Stage & 6-Stage & 8-Stage & 10-Stage \\
    \hline\hline
    LeNet-5 & 5 & (1) & (1,2) & (1,2,3) & (1,2,3,4) \\
    \hline
    AlexNet & 8 & (1) & (1,2) & (1,2,3) & N/A\\
    \hline
    VGG-16 & 16 & (2) & (2,4) & (2,4,7) & (2,4,7,10) \\
    \hline
    ResNet-20 & 20 & (7) & (7,13) & (7,13,19) & N/A\\
    \hline
    \end{tabular}
    \end{center}
    \vspace{-0.1in}
    \caption{Pipeline Placement Vectors for CNNs}
    \label{table:pipeline_schemes}
    \end{table*}

Table~\ref{table:simulated_pipeline_accuracy} shows the inference 
accuracy obtained after up to 30,000 iterations of training. For 
LeNet-5, the inference accuracy drop is within 0.5\%. However, for 
the other networks, there is a small drop in inference accuracy 
with 4 and 6 stages. AlexNet has about 4\% drop in inference 
accuracy, but for VGG-16 the inference accuracy drop is within 
2.4\%, and for ResNet-20 the accuracy drop is within 3.5\%. Thus, 
the resulting model quality is comparable to that of a 
non-pipelining-trained model. 

However, with deeper pipelining (i.e., 8 and 10 stages), inference 
accuracy significantly drops. There is a 12\% and a 8.5\% inference 
accuracy drop for VGG-16 and ResNet-20 respectively. 
In this case, the model quality is not comparable to that of 
the non-pipelined training. This results confirm what is reported
in the literature~\citep{PipeDream} and can be attributed to the 
use of stale weights. Below we further explore the impact of stale 
weights on inference accuracy.

\begin{table*}[t]
    \begin{center}
    \begin{tabular}{| c | c | c | c | c | c |}
    \hline
    CNN & Non-pipelined & 4-Stage & 6-Stage & 8-Stage & 10-Stage \\
    \hline\hline
    LeNet-5 & 99.00\% & 98.64\% & 98.62\% & 98.61\% & 98.47\% \\
    \hline
    AlexNet & 82.51\% & 78.47\% & 78.32\% & 78.47\% & N/A\\
    \hline
    VGG-16 & 91.36\% & 90.53\% & 88.96\% & 83.73\% & 79.85\% \\
    \hline
    ResNet-20 & 91.50\% & 90.05\% & 88.00\% & 83.01\% & N/A\\
    \hline

    \end{tabular}
    \end{center}
     \vspace*{-0.1in}
    \caption{Inference Accuracy for Simulated Pipelined Training}
    \label{table:simulated_pipeline_accuracy}
    \end{table*}

\subsection{Impact of Weight Staleness}

We wish to better understand the impact of the number of pipeline 
stages and the location of these stages in the network on inference accuracy. 
We focus on ResNet-20 because of its relatively small size and regular 
structure. It consists of 3 residual function groups with 3 residual 
function blocks within each group. In spite of this relatively small size 
and regular structure, it enables us to create pipelines with up to 20 
stages by inserting pipeline register pairs within residual function 
blocks.

\begin{figure}[t]
\centering
\begin{minipage}{0.5\textwidth}
    \centering
     \vspace*{-0.1in}
    \captionsetup{type=table} 
    \begin{tabular}{| c | c | c |}
    \hline
    Stages        & Inference Accuracy\\
    \hline\hline
    Non-pipelined & 91.50\% \\
    \hline
    8 & 90.28\% \\
    \hline
    10 & 88.37\% \\
    \hline
    12 & 88.73\% \\
    \hline
    14 & 87.94\% \\
    \hline
    16 & 87.30\% \\
    \hline
    18 & 86.23\% \\
    \hline
    20 & 79.09\% \\
    \hline
    \end{tabular}
    \caption{Fine-grained Pipelining Inference Accuracy}
    \label{table:Fine-grained Pipelined ResNet-20}
\end{minipage}%
\begin{minipage}{0.5\textwidth}
    \centering
    \includegraphics[width=2.5in]{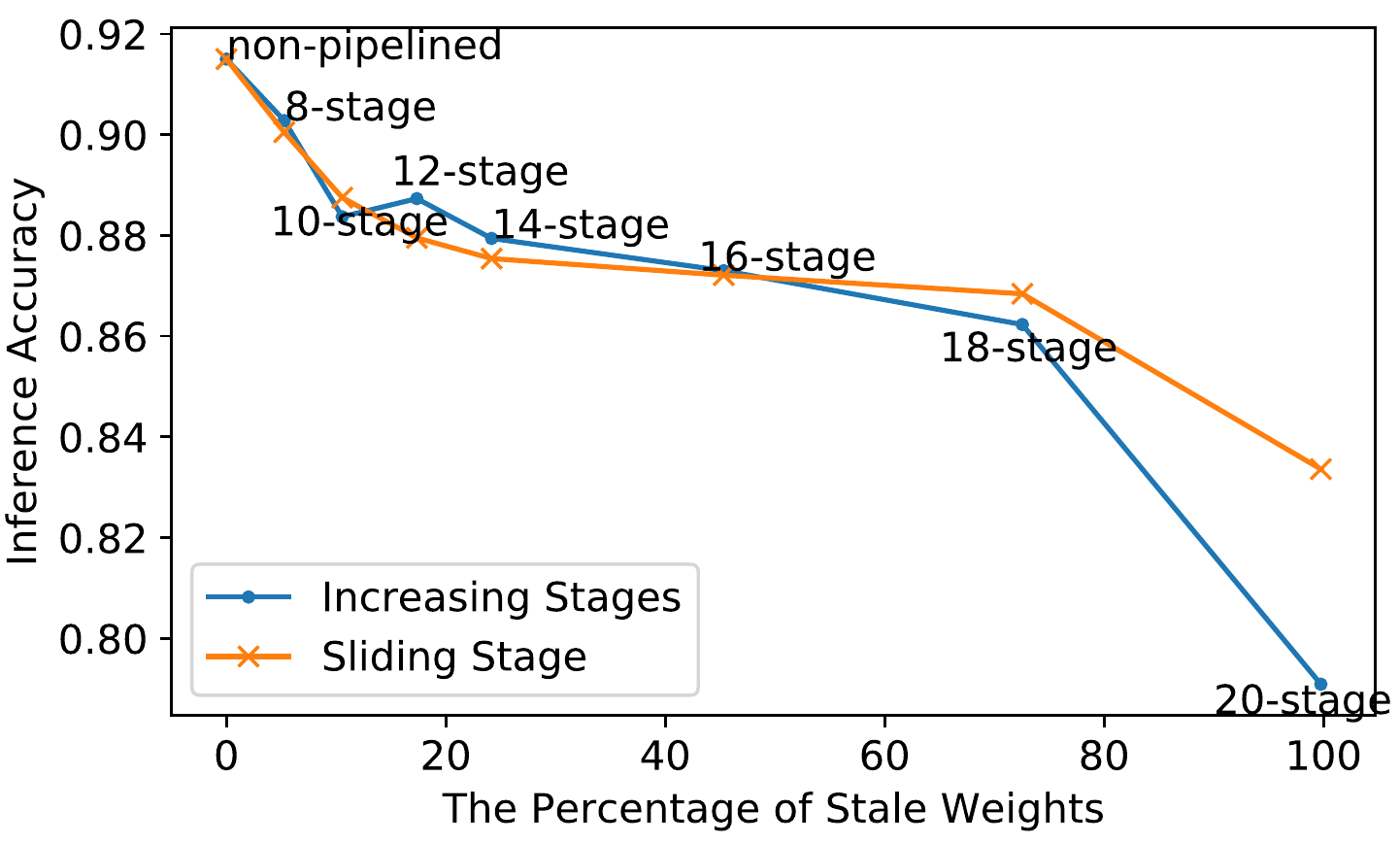}
     \vspace*{-0.1in}
    \caption{Inference Accuracy vs \% of Stale Weights}
    \label{fig:accuracy-stale}
\end{minipage}
\end{figure}

We conduct two experiments. In the first, we increase the number of 
pipeline stages (from earlier layers to latter layers) and measure
the inference accuracy of the resulting model. The results are shown in 
Table~\ref{table:Fine-grained Pipelined ResNet-20}, which gives the 
inference accuracy of pipelined training after 100,000 iterations, 
as the number of pipeline stages increases. The 8-stage pipelined training
is created by a $\rm{PPV}$ of (3,5,7), and the subsequent pipeline schemes are 
created by adding pipeline registers after every 2 layers after layer 7. 
Clearly, the greater the number stages, the worse is the resulting model quality.

The number of stale weights used in the pipelined training increases as 
the number of pipeline stages increases. Thus, Figure~\ref{fig:accuracy-stale} 
depicts the inference accuracy as a function of the percentage of weights that
are stale. The curve labeled ``Increasing Stages'' shows that the drop in 
inference accuracy increases as the percentage of stale weights increases. 


In the second experiment, we investigate the impact of the {\em degree
of staleness} (Section~\ref{sec:pipelined}). 
Only {\em one} pair of pipeline 
registers is inserted. The position of this register slides from the 
beginning of the network to its end. At every position, the percentage 
of stale weights remains the same as in the first experiment, but all 
stale weights have the same degree of staleness.  The result of this 
experiment is shown by the curve labeled ``Sliding Stage'' in 
Figure~\ref{fig:accuracy-stale}. The curve shows the inference accuracy also
drops as the percentage of stale weights increases. However, it also
indicates that the drop of inference accuracy remains more or less the 
same as in the first experiment in which the degree of staleness is higher. 
Thus, the percentage of stale weight appears to be what determines the 
drop in inference accuracy and not the degree of staleness of the weights. 

The percentage of stale weight is determined by where the last 
pair of pipeline registers are placed in the network. It is the position
of this pair that determines the loss in inference accuracy. Therefore,
it is desirable to place this last pair of registers as early as 
possible in the network so as to minimize the drop in inference 
accuracy.

While at first glance this may seem to limit pipelining, it is important
to note that the bulk of computations in a CNN is in the first few 
convolutional layers in the network. Inserting pipeline registers for
these early layers can result in both a large number of stages that are
computationally balanced. For example, our profiling of the runtime of 
ResNet-20 shows that the first three residual functions take more than 
50\% of the training runtime.  This favors more pipeline stages at 
the beginning of the network. Such placement has the desirable effect 
of reducing the drop in inference accuracy while obtaining relatively 
computationally balanced pipeline stages.

\subsection{Effectiveness of Hybrid Training}

\begin{figure}[t]
\centering
\begin{minipage}{0.5\textwidth}
    \centering
    \includegraphics[height=0.85in]{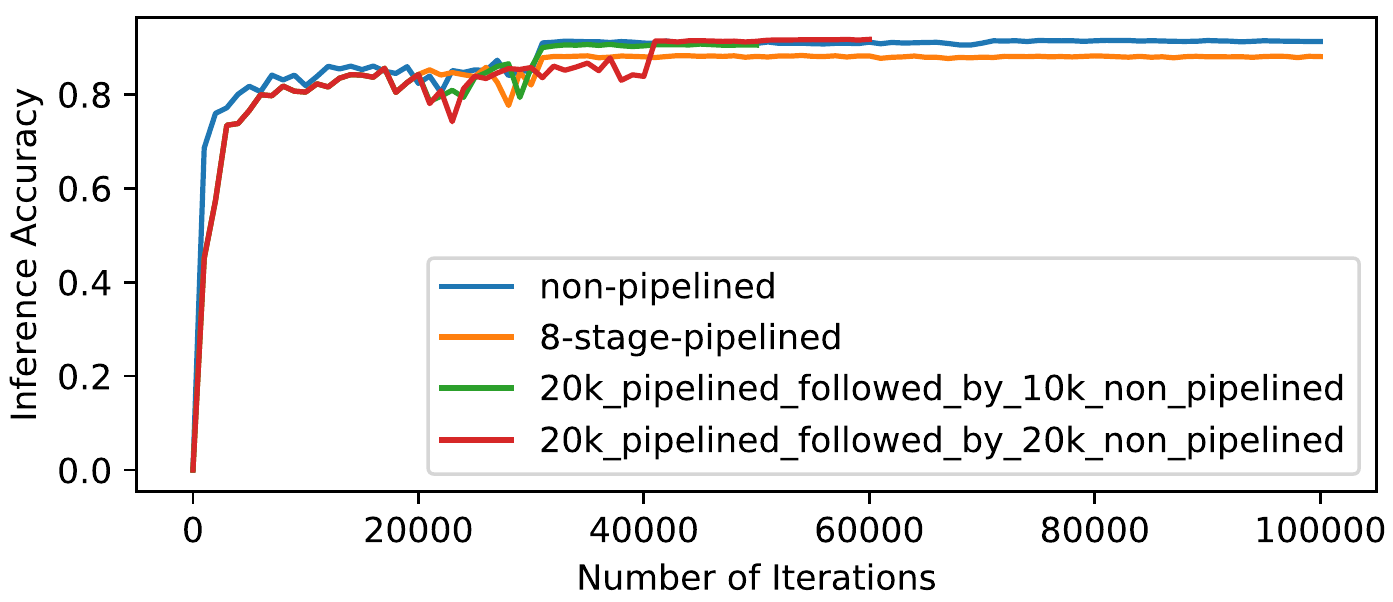}
     \vspace*{-0.1in}
    \caption{Hybrid Training Convergence}
    \label{fig:ResNet20-hybrid}
\end{minipage}%
\begin{minipage}{0.5\textwidth}
\centering
     \vspace*{-0.1in}
    \captionsetup{type=table} 
    \begin{tabular}{| c | c | c |}
    \hline
                 & Inference Accuracy \\
    \hline\hline
    Baseline 30k & 91.50\% \\
    \hline
    Pipelined 30k & 88.29\% \\
    \hline
    20k+10k hybrid & 90.71\% \\
    \hline
    20k+20k hybrid & 91.72\% \\
    \hline
    \end{tabular}
    \caption{Hybrid Training Inference Accuracy}
    \label{table:ResNet20-hybrid}
\end{minipage}
\end{figure}

We demonstrate the effectiveness of hybrid training using only ResNet-20 for 
brevity. Figure~\ref{fig:ResNet20-hybrid} shows the inference accuracy 
for 20K iterations of pipelined training followed by either 10K or 20K 
iterations of non-pipelined training. This inference accuracy is 
compared to 30K iterations of either non-pipelined or pipelined training
with $\rm{PPV}$ (5,12,17).
The figure demonstrates that hybrid training converges in a similar manner
to both pipelined and non-pipelined training. Table~\ref{table:ResNet20-hybrid} 
shows the resulting inference accuracies. The table shows that the 20K+10K
hybrid training produces a model with accuracy that is comparable to that
of the non-pipelined model. Further, with an additional 10K iterations
of non-pipelined training, the model quality is slightly better than that
of the non-pipelined model. This demonstrates the effectiveness of 
hybrid training.

\subsection{Pipelined and Hybrid Training Performance}

We implement 4-stage pipelined training ResNet-20/56/110/224/362
on a 2-GPU system. Each GPU is responsible for one forward stage and one backward 
stage. Thus, the maximum speedup that can be obtained is 2. We train every
ResNet for 200 epochs.  Table~\ref{table:performance results} 
shows the inference accuracies with and without pipelining as model as the 
measured speedups of pipelined training over the non-pipelined one. The table 
indicates that the quality of the models produced by pipelined training is 
comparable to those achieved by the simulated pipelining on {\em Caffe}. The table 
further shows that speedup exists for all networks. Indeed, for ResNet-362, the 
speedup is 1.82X. This is equivalent to about 90\% utilization for each GPU. The 
table also reflects that as the networks get larger, the speedup improves. This 
reflects that with larger networks, the ratio of computation to communication 
overhead is higher, leading to better speedups. 

Moreover, we combine the 4-stage pipelined training described above with non-pipelined
training to demonstrate the performance of hybrid training. We train every ResNet using 
pipelined training for 100 epochs and follow it up by 100 epochs of non-pipelined training.
Because the maximum speedup for the pipelined training is 2 and only half the training epochs 
is accelerated, the maximum speedup for this hybrid training is $s = t/(t/2 + t/4) = 1.33$, where
$t$ is the training time of non-pipelined training. Table~\ref{table:performance results} shows 
the inference accuracies and speedup of the hybrid training for each ResNet and validates that 
hybrid training can produce a model quality that is comparable to the baseline non-pipelined 
training while speeding up the training process. As network size grows, the speedup reaches 1.29X,
approaching the theoretical limit 1.33X. 

\begin{table*}[t]
    \begin{center}
    \resizebox{\textwidth}{!}{%
    \begin{tabular}{| c | c | c | c | c | c | c | c | c | c |}
    \hline
    & & \multicolumn{3}{|c|}{Accuracy}  & \multicolumn{3}{|c|}{Time(Seconds)}& \multicolumn{2}{|c|}{Speedup}\\
    \cline{3-10}
    \multirow{2}{*}{ResNet} & \multirow{2}{*}{PPV} & Non- & \multirow{2}{*}{Pipelined} & \multirow{2}{*}{Hybrid}& Non-& \multirow{2}{*}{Pipelined} & \multirow{2}{*}{Hybrid} & \multirow{2}{*}{Pipelined} & \multirow{2}{*}{Hybrid}\\
    & & pipelined & & & pipelined & & & &  \\
    \hline
    -20 & (7) & 91.65\% & 91.21\% & 91.37\% & 2,421 & 1,974 & 2,205 & 1.23X & 1.10X\\
    \hline
    -56 & (19) & 92.63\% & 92.89\% & 92.75\% & 6,745 & 4,090 & 5,429 & 1.65X & 1.24X\\
    \hline
    -110 & (37) & 93.59\% & 92.88\% & 93.55\% & 13,150 & 7,570 & 10,452 & 1.73X & 1.26X\\
    \hline
    -224 & (75) & 92.77\% & 91.39\% & 93.33\% & 27,231 & 14,998 & 21,245 & 1.81X & 1.28X\\
    \hline
    -362 & (121) & 93.46\% & 90.53\% & 93.98\% & 44,814 & 24,640 & 34,814 & 1.82X & 1.29X\\
    \hline
    \end{tabular}}
    \end{center}
     \vspace*{-0.1in}
    \caption{Inference Accuracy and Speedup of Actual Pipelined and Hybrid Training}
    \label{table:performance results}
    \end{table*}

\subsection{Memory Usage}

Pipelined training requires the saving of intermediate activations, as described 
earlier in Section~\ref{sec:pipelined}, leading to an increase in memory footprint.
This increase in memory is a function of not only the placement of the pipeline
registers, but also of the network architecture and the number of inputs in a 
mini-batch (batch size). We calculate the memory usage of the 4-stage pipelined 
ResNet training above to show that this increase is modest for our pipelining 
scheme. Specifically, we use  {\em torchsummary} in {\em PyTorch} to report memory 
usage for weights and activations for a network and calculate the additional memory 
required by the additional copies of activations. The results are shown in 
Table~\ref{tab:memory usage}. Assuming a batch size of 128, the percentage increase
in size is close to 60\% except for ResNet-20.

\begin{table*}[t]
    \begin{center}
    \begin{tabular}{| c | c | c | c | c | c |}
    \hline
     ResNet & PPV & Activations & Weight & Increase & Increase \% \\
    \hline
    -20 & (7) & 3.84MB x batch size & 1.03MB & 2.58MB x batch size & 67\% \\
    \hline
    -56 & (19) & 10.87MB x batch size & 3.25MB & 6.32MB x batch size & 58\% \\
    \hline
    -110 & (37) & 21.43MB x batch size & 6.59MB & 12.35MB x batch size & 57\% \\
    \hline
    -224 & (75) & 43.70MB x batch size & 13.64MB & 25.07MB x batch size & 57\% \\
    \hline 
    -362 & (121) & 70.67MB x batch size & 22.17MB & 40.50MB x batch size & 57\% \\
    \hline
    \end{tabular}
    \end{center}
     \vspace*{-0.1in}
    \caption{Memory Usage of 4-Stage Pipelined ResNet Training}
    \label{tab:memory usage}
    \end{table*}

\subsection{Comparison to Existing Work}
\label{sec:comparison}


We compare our pipelined training scheme with two key existing systems: PipeDream~\citep{PipeDream}
and GPipe~\citep{GPipe}. We do so on three aspects: the pipelining scheme, performance and memory
usage. We notice that some aspects of previous work in literature by Huo et al.~\citep{Huo_Decoupled,Huo_FR}, 
Decoupled Backpropagation (DDG) and Feature Replay (FR) are similar to PipeDream and GPipe respectively while 
providing a detailed and rigorous convergence analysis. We discuss DDG and FR in Section~\ref{sec:related}.  

Our pipelining scheme is simpler than that of PipeDream and GPipe in that we do not
require weight stashing nor do we divide mini-batches into micro-batches. This leads to less 
communication overhead, and is amicable to rapid realization 
in machine learning framework such as {\em PyTorch} or in actual hardware such as Xilinx's xDNN
FPGA accelerators~\citep{xDNN}.

Our pipelining scheme, as PipeDream, eliminates bubbles that exist in the pipeline leading to
better performance. For example, we obtain a speedup of 1.7X for ResNet-110 using 2 GPUs in contrast
to GPipe that obtains a speedup of roughly 1.3X for ResNet-101 using 2 TPUs. We also obtain similar performance 
compared to PipeDream for similar networks. When the number of pipeline stages grows, pipeline bubbles 
exhibits more negative effect on performance shown in GPipe on a 4-partition pipelined ResNet-101 using 4 TPUs as
its bubble overhead doubled compared to that of the 2-partition pipelined ResNet-101.

Our scheme uses less memory compared to PipeDream, although it introduces more memory overhead 
compared to GPipe. PipeDream saves intermediate activations during training, as we do. However, 
it also saves multiple copies of a network's weights for weight stashing.
The memory footprint increase due to this weight stashing depends on the network architecture,
including the number of weights and activations, as well as on the size of the mini-batch. 
For example, for VGG-16 trained on CIFAR-10 with a mini-batch size of 128 using a 4-stage,
pipelined training, we estimate our pipelining methodology to use 49\% less memory compared 
PipeDream. Similarly for VGG-16 trained on ImageNet~\citep{imagenet_cvpr09} and a mini-batch size of 32,
our scheme uses 29\% less memory.
We estimate the memory increase due to weight stashing also using {\em tourchsummary}.

\section{Related Work}
\label{sec:related}

There has been considerable work that explores parallelism in the 
training of deep neural networks. There are several approaches to exploiting 
parallelism. 

One approach is to exploit {\em data 
parallelism}~\citep{ChenMBJ16,Cui16,GoyalDGNWKTJH17,Poseidon,Dean_et_al, Wang2019BlinkFA},
in which each accelerator obtains a full copy of the model and processes different 
mini-batches of training data simultaneously. At the end of each training iteration, 
the gradients produced by all accelerators are aggregated and used to update weights 
for all copies of the model, synchronously~\citep{ChenMBJ16,GoyalDGNWKTJH17} or 
asynchronously~\citep{Dean_et_al}. A centralized parameter server is usually used 
to facilitate data communication~\citep{Cui16,Dean_et_al}.  Although the training 
is performed in parallel, the communication overhead can be 
significant~\citep{Wang2019BlinkFA}.

A second approach is to exploit 
{\em model parallelism}~\citep{Kim,LeeNIPS2014,Project_Adam,Dean_et_al,GraphLab}. 
In this approach, a model is partitioned onto different 
accelerators~\citep{Kim,LeeNIPS2014,GraphLab, Project_Adam, Dean_et_al}. 
Each accelerator is only responsible for updating the weights for the portion of 
the model assigned to it. This approach is often used when a model is large and cannot 
fit into the memory of a single accelerator.
However, because of the data dependences described in Section~\ref{sec:backprop}, 
only one accelerator is active during the training process, resulting in 
under-utilization of accelerators resources. Moreover, inter-layer activations and 
gradients across two consecutive stages needs to be communicated during training, 
adding more overhead to the entire process.

{\em Pipelined parallelism} addresses the under-utilization of accelerators resources for 
the training of large models.  There have been a few studies that explore pipelined 
parallelism~\citep{petrowski,Chen2012PipelinedBF,Mostafa,PipeDream,GPipe,Huo_Decoupled,Huo_FR},
which we review in this section.

PipeDream~\citep{PipeDream} implements pipelined training for large neural networks such
as VGG-16, Inception-v3 and S2VT across multiple GPUs. 
However, in their implementation, they limited the usage of stale weights by 
{\em weight stashing}, i.e., keeping multiple versions of network parameters (weights) 
during training. This increases the memory footprint of training. In contrast, we do 
not maintain multiple copies of weights during training, therefore reducing the memory 
footprint of pipelined training. 

GPipe~\citep{GPipe} implements a library in \emph{Tensorflow} to enable pipelined 
parallelism for the training of large neural networks.  GPipe pipelines micro-batches 
within each mini-batch to keep the gradients consistently accumulated. This eliminates
the use of stale weight during training, but it does so at the expense of ``pipeline 
bubbles'' at steady state. GPipe utilizes these bubbles to reduce the 
memory footprint by re-computing forward activations instead of storing them.
In contrast, our work has no pipeline bubble and thus dedicates computing resources 
to compute forward pass and backward pass only once during each training iteration. 

Huo et al.~\citep{Huo_Decoupled} implement decoupled backpropagation (DDG) using 
delayed gradient updates. They show that DDG guarantees convergence through a 
rigorous convergence analysis. Similar to PipeDream, DDG uses multiple copies of
the weights and thus increases memory footprint. Further, DDG pipelines only 
the backward pass of training, leaving forward pass un-pipelined.
Huo et al.~\citep{Huo_FR} follow up by proposing feature replay (FR) that 
re-computes activations during backward pass, similar to GPipe, resulting less 
memory footprint and improved inference accuracy than DDG. 
In contrast, we pipeline both forward and backward pass without maintaining 
multiple copies of weights or re-computing forward activations during backward 
pass.   

Thus, in summary, our work contrasts to the above work on pipelined training, 
in that we use pipelining with unconstrained stale weights, resulting in full 
pipeline utilization with a modest increase in memory usage. We extend earlier 
work by studying the impact of weights staleness on the quality of the model. 
We show that it is effective 
to use stale weights if the pipelining is in early layers, which is where 
the bulk of computations exist. Further we also extend earlier work through 
hybrid training, which combines both pipelined and non-pipelined training.
We compare the performance and memory footprint increase of our scheme to
existing work in Section~\ref{sec:comparison}.

\section{Concluding Remarks}
\label{sec:conclusions}

We evaluate pipelined execution of backpropagation for the training of CNNs
in a way that fully utilizes accelerators, achieving a speedup of 1.82X on
the 2-GPU system, and does not significantly increase memory usage, unlike previous work. 
We show that pipelining training with stale weights does converge. 
Further, we show that the inference accuracies 
of the resulting models are comparable to those of models obtained with 
traditional backpropagation, but only when pipelining is implemented in
the early layers of the network, with inference accuracy drop within 1.45\% 
on 4-stage pipelined training except for AlexNet. This does not limit the benefit of 
pipelining since the bulk of computations is in the early convolutional 
layers. When pipelining is implemented deeper in the network, the
inference accuracies do drop significantly, but we can compensate for this drop by
combining pipelined with non-pipelined training, albeit with lower 
performance gains, obtaining model quality with an average of 0.19\% better than 
the baseline in inference accuracies for ResNets.

This work can be extended in a number of directions. One direction
is to evaluate the approach with a larger number of
accelerators since pipelined parallelism is known to scale naturally 
with the number of accelerators. Another is to evaluate the approach on
larger datasets, such as ImageNet. Finally, pipelined parallelism lends 
itself to hardware implementation. Thus, another direction for future 
work is to evaluate pipelined parallelism using Field Programmable 
Gate Array (FPGA) or ASIC accelerators.


\setspacing{0.90}
\newpage
\bibliography{paper}
\bibliographystyle{iclr2020_conference}

\newpage
\appendix
\label{app:hyper}
\section{Training Hyperparameters for Simulated Pipelined Training}
LeNet-5 is trained on the MNIST dataset with Stochastic Gradient Descent (SGD) 
using a learning rate of 0.01 with inverse learning policy, a momentum of 0.9, 
a weight decay of 0.0005 and a mini-batch size of 100 and for 30,000 iterations. 
The progression of inference accuracy during training is recorded with 300 tests 

AlexNet is trained on the CIFAR-10 dataset with SGD with Nesterov momentum using 
a learning rate of 0.001 that is decreased by 10x twice during training, a 
momentum of 0.9, a weight decay of 0.004 and a mini-batch size of 100 for 250,000 
iterations. One test is performed every epoch to record the progression of 
inference accuracy.

VGG-16 is trained on CIFAR-10 dataset with SGD with Nesterov momentum using a 
learning rate starting at 0.1 that is decreased by half every 50 epochs during 
training, a momentum of 0.9, a weight decay of 0.0005 and a mini-batch size of 
100 for 250,000. Since it is relatively more difficult to train VGG-16 compared 
to other models,  batch normalization and dropout are used during training 
throughout the network. One test is performed every epoch to record the 
progression of inference accuracy.

ResNet is trained on CIFAR-10 dataset with SGD using a learning rate starting 
at 0.1 and 0.01 for non-pipelined and pipelined training respectively, that 
is decreased by 10x twice during training, a momentum of 0.9, a weight decay 
of 0.0001 and a mini-batch size of 128 for 100,000 iterations. Batch 
normalization is used during training throughout the network. One test is 
performed every 100 iterations to record the progression of inference accuracy.

\section{Training Hyperparameters for Actual Pipelined Training}
For the baseline non-piplined training, ResNet-20/56/110/224/362 is trained on CIFAR-10 
dataset for 200 epochs with SGD using a learning rate of 0.1 that is decreased by 
a factor of 10 twice (at epoch 100 and 150), a momentum of
0.9, a weight decay of 0.0001 and a mini-batch size of 128. Batch 
normalization is used during training throughout the network. This set of hyperparameters
can be found at https://github.com/akamaster/pytorch\_resnet\_cifar10. 

For the 4-stage pipelined training, the hyperparameters are the same as the non-pipelined baseline,
except for the ${\rm BKS}_{2}$ learning rate. Table~\ref{tab:learning_rate_part1}
shows that learning rate for all ResNet experimented.

\begin{table*}[h]
    \begin{center}
    \begin{tabular}{| c | c |}
    \hline
           & ${\rm BKS}_{2}$ learning rate  \\
    \hline
    ResNet-20 &  0.1 \\
    \hline
    ResNet-56 &  0.01 \\
    \hline
    ResNet-110 & 0.001 \\
    \hline
    ResNet-224 & 0.001 \\
    \hline 
    ResNet-362 & 0.001 \\
    \hline
    \end{tabular}
    \end{center}
     \vspace*{-0.1in}
    \caption{Learning Rate of ${\rm BKS}_{2}$}
    \label{tab:learning_rate_part1}
    \end{table*}

\end{document}